\documentclass[preprint,aps,floatfix,showpacs,a4paper]{revtex4-1}

\usepackage{amsmath,amssymb,amsfonts,dcolumn,eulervm,color,graphicx,graphics,latexsym,placeins,epsfig,subfigure,hyperref}

\newcommand{\be}{\begin{equation}}

\newcommand{\ee}{\end{equation}}

\newcommand{\lb}{\left}

\newcommand{\rb}{\right}

\newcommand{\cG}{\mathcal{G}} 

\newcommand{\cL}{\mathcal{L}} 

\newcommand{\cE}{\mathcal{E}} 

\newcommand{\cF}{\mathcal{F}} 

\newcommand{\wg}{\wedge}    

\newcommand{\arxiv}[1]{\href{http://arxiv.org/abs/#1}{arXiv:#1}}    

\definecolor{darkred}{rgb}{.8,0,0}

\definecolor{darkblu}{rgb}{0,0,.8}

\definecolor{darkgreen}{rgb}{0,.8,0}

\begin{document}


\title{On the static Lovelock black holes }

\author{Naresh Dadhich}\affiliation{ Inter-University Centre for Astronomy \& Astrophysics,\\ Post Bag 4, Pune 411 007, India}\email{nkd@iucaa.ernet.in}

\author{Josep M. Pons}\affiliation{ DECM and ICC, Facultat de F\'{\i}sica, Universitat de Barcelona,\\ Diagonal 647, 08028 Barcelona, Catalonia, Spain.}\email{pons@ecm.ub.es}

\author{Kartik Prabhu}\affiliation{Department of Physics, University of Chicago,\\5640 S. Ellis Avenue, Chicago, IL 60637, USA}\email{kartikp@uchicago.edu}

\begin{abstract}

We consider static spherically symmetric Lovelock black holes and generalize the dimensionally continued black holes in such a way that they
asymptotically for large $r$ go over to the d-dimensional Schwarzschild black hole in dS/AdS spacetime. This means that the master algebraic polynomial is not
degenerate but instead its derivative is degenerate. This family of solutions contains an interesting class of pure Lovelock black holes which
are the $N$th order Lovelock $\Lambda$-vacuum solutions having the remarkable property that their thermodynamical parameters have the universal
character in terms of the event horizon radius. This is in fact a characterizing property of pure Lovelock theories. We also demonstrate the universality of the asymptotic Einstein limit for the Lovelock black holes in general.
\end{abstract}


\maketitle

\section{Introduction}

With a view to understand gravity in higher dimensions, there has been extensive work on various generalizations of General Relativity in higher
dimensional spacetimes. The modifications should however be consistent with the following general features: (a) general covariance - the
Lagrangian must be a scalar density constructed from the Riemann curvature which yields a non-trivial equation of motion, (b) the equivalence
principle and (c) the equation of motion to be second order quasi-linear. This uniquely identifies the Lanczos-Lovelock Lagrangian (LL-gravity)
which is a homogeneous polynomial in the Riemann curvature with specific coefficients where zeroth, linear and quadratic orders respectively
correspond to the cosmological constant, Einstein-Hilbert and Gauss-Bonnet terms \cite{lovelock}. It is pertinent here to note that the
Gauss-Bonnet term also arises in the one loop correction of the low-energy effective action in string theory \cite{string}. Not only string
theory but to physically realize high energy effects of gravity within the classical framework also asks for higher dimensions and the inclusion
of such higher order terms in the gravitational action. It is important that this is a purely classical motivation for higher dimensions for the
consideration of high energy effects \cite{dad}. The requirement (c) ensures the unique physical evolution for a given initial value problem.
Further it is interesting that the Lanczos-Lovelock Lagrangian is also characterized by several different considerations which include the
derivation of the equation of motion by the Bianchi derivative \cite{Dadhich:2008df} and the equivalence of the metric and Palatini formulation
for arbitrary connection by the Levi-Civita consistent truncation \cite{Dadhich:2010}. The LL-gravity is therefore the most natural
generalization of the Einstein gravity in the strong gravitational regime where higher order curvature terms may become important and represent high
energy corrections. \\

In this work, we shall employ the study of properties of the static spherically symmetric Lovelock black holes for understanding gravitational
dynamics in higher dimensions. Spherically symmetric black hole solutions were first discovered for Gauss-Bonnet extension of GR in five
dimensions \cite{boul deser, wiltshire}. There is a very extensive body of work on black holes in general Lovelock theories beginning with the
three classic papers \cite{wheeler,whitt,MS} and followed by \cite{LovelockBH}. For static spherically symmetric vacuum solutions, the equation
ultimately reduces to an algebraic equation involving an $N$-th order polynomial (henceforth called the master equation/polynomial). The problem
then reduces to simply solving this polynomial. As is well known there exists no standard method to solve it for $N>4$. One of the obvious
strategies could be to assume that the polynomial is degenerate so that it is trivially solved. This is precisely what is done for the
dimensionally continued black hole solutions \cite{BTZ,CTZ}. It has been motivated by the considerations of the extension of the Euler density
to the next higher dimension as well as the embedding of Lorentz group $SO(d-1,1)$ into the larger AdS group $SO(d-1,2)$ and above all a unique
value of $\Lambda$. This would mean that all Lovelock coefficients are not independent but are given in terms of the single one, $\Lambda$. Thus
dimensional continuity prescription is nothing but the complete degeneracy of the master polynomial. This is all very fine but the solution does
not asymptotically go over to the corresponding Einstein solution for large $r$. By the corresponding Einstein solution, we mean the $d$-dimensional Schwarzschild black hole in dS/AdS spacetime. Henceforth this is what would be implied by the Einstein solution or limit. As the higher order Lovelock terms are supposed to give
correction to the Einstein gravity, it is therefore pertinent that the solution should have the right Einstein limit far away from the source. \\

The question is, could the assumption of complete degeneracy be modified such that the solution asymptotically approaches the Einstein solution?
The natural generalization of degeneracy is the derivative degeneracy; i.e. the master polynomial is not degenerate but its first derivative is
completely degenerate. That is, it could be written as $(X - \beta)^N -\alpha$ which is also trivially solvable. The solution would then tend to
the degeneracy case of dimensionally continued black holes \cite{BTZ,CTZ} for $r\to r_h$ where $r_h$ is the horizon radius while for large $r$
it would tend to the corresponding Einstein solution. It reduces to the degeneracy case for $\alpha=0$ and for $\beta = 0$ it is the pure $N$-th
order Lovelock black hole solution with $\Lambda$. It is the solution of the pure Lovelock vacuum equation with $\Lambda$ and we would
henceforth simply call it pure Lovelock black hole \cite{dad pure L}. The driving consideration for the dimensionally continued black holes was
to have all the couplings given in terms of the unique value of $\Lambda$ so that the thermodynamical information could be easily extracted. The
pure Lovelock black hole also has only one parameter, $\alpha$ and hence is as benign and interesting as the dimensionally continued one. In
addition it has the desired asymptotic limit as it goes over to the corresponding Einstein solution for large $r$. \\

Besides this, the pure Lovelock black hole has the remarkable characterizing property that its thermodynamical parameters bear an universal relation to the horizon radius in the critical $d=2N+1, 2(N+2)$ dimensions\cite{kpd-lett}. That is the thermodynamical parameters, temperature and entropy bear the same scaling relation with the horizon radius $r_h$ for odd ($d=2N+1$)
and even ($d=2(N+1)$) dimensions irrespective of the Lovelock order $N$. For instance, the entropy would always scale as $r_h$ and $r_h^2$
respectively for odd and even dimensions. The thermodynamics is therefore entirely insensitive or neutral to the Lovelock order. Not only this,
the converse is also true. The universality of thermodynamics uniquely characterizes the pure Lovelock black hole. This is in line with the
first universal feature of higher dimensional gravity discovered for the uniform density fluid sphere \cite{dmk}. The Schwarzschild interior
solution always describes the uniform density sphere for the Einstein as well as the Einstein-Lovelock gravity. 

Note that $d=2N+1, 2(N+2)$ are the critical dimensions for the $N$th order Lovelock gravity in the same sense as $d=3,4$ are for the Einstein gravity. The vacuum solution in $3$ dimension is trivially flat and it becomes non-trivial in $4$ dimension. This is true in general for $d=2N+1, 2(N+2)$ in general for the $N$th order analogue of the Riemann curvature (i.e. the $N$th order vacuum is trivial in $d=2N+1$ dimension). This is what has recently been shown \cite{odd} and it has been motivated by the universality of thermodynamics of the pure Lovelock black holes. Thus $d=2N+1, 2(N+2)$ are the critical dimensions having the similar behavior for the $N$th order Lovelock gravity.\\

Further we would also show that all vacuum solutions with $\Lambda$ have the
universal asymptotic behavior tending to the corresponding Einstein limit
irrespective of whether it is pure Lovelock or Einstein-Lovelock (summing
over all Lovelock $\alpha_i$ ) solution. In the latter case if there are repeated roots of the master polynomial (which means all $\alpha_i$'s are not independent), there won't exist the asymptotic Einstein limit for that root. The non-degenerate character of the polynomial is necessary for the asymptotic Einstein limit. In the case of pure Lovelock, the polynomial is required to be non-degenerate ensuring the proper Einstein limit.

The paper is organised as follows: Sec.\ref{sec:LL} summarizes the LL theory which is followed in Sec.\ref{sec:BH} by the discussion of the
static Lovelock black holes. Next we consider the thermodynamical universality in Sec.\ref{sec:therm} followed by the asymptotic limit of the
solutions in Sec.\ref{sec:asym}. We conclude with a discussion. \\[15pt]

\section{Lanczos-Lovelock gravity}\label{sec:LL}

    Consider the $d$-dimensional spacetime to be equipped with $1$-form vielbeins $e^a$ (such that the metric $g=\eta_{ab}~e^a\otimes e^b$) and a local $1$-form spin connection $\omega^a_{~b}$. The torsion $T^a$ and curvature $R^a_{~b}$ are defined as:
    \begin{subequations}\label{eq:defn T R}
    \begin{align}
         T^a & = de^a + \omega^a_{~b} \wg e^b \label{eq:defn T} \\
        R^a_{~b} & = d\omega^a_{~b} + \omega^a_{~c} \wg \omega^c_{~b}
    \end{align}
    \end{subequations}
where $a,b\ldots$ represent the Lorentz indices in a local orthonormal frame. In the first order formalism, the general LL-gravity Lagrangian density can be simply written as:
    \be\label{eq:lagrangian}
        \cL = \frac{1}{2\kappa}\sum_{k=0}^{N}\bar\alpha_k \cL^{(k)} =  \frac{1}{2\kappa}\sum_{k=0}^{N}\frac{\alpha_k}{(d-2k)(d-2)!} \cL^{(k)}
    \ee
where $\alpha_k$ are the couplings for the various terms, and $\cL^{(k)}$ denotes the $k^{th}$ order polynomial composed of dimensionally
continued Euler densities:
    \be\label{eq:lovelock poly}
        \cL^{(k)} = \epsilon_{a_1\ldots a_d}~R^{a_1a_2}\wg \cdots \wg R^{a_{2k-1}a_{2k}} \wg e^{a_{2k+1}} \wg \cdots \wg e^{a_d}
    \ee
The maximum order polynomial that contributes in $d$-dimensions to the field equations is given by:
    \be\label{eq:N defn}
        N = \lb\lfloor\frac{d-1}{2}\rb\rfloor
    \ee
Any term of order greater than $N$ is  either zero or at best a topological invariant and hence does not contribute to the classical field equations. In what follows the limits on the sum are omitted, and assumed from $0$ to $N$ unless otherwise stated.\\

Written with only spacetime indices, we have the Lagragian $L^{(0)} \sim -2\Lambda$;  $L^{(1)} \sim R$, the Einstein-Hilbert and the next quadratic
Gauss-Bonnet, $L^{(2)} \sim \cG = R^2 - 4R_{ij} R^{ij} + R_{ijkl} R^{ijkl}$.  Restricting to the case of vanishing torsion $T^a =0$, the
variation of the action with respect to $\omega^a_{~b}$ identically vanishes and variation with $e^a$ yields the field equations of the form:
        \be\label{eq:field eqns}
            \sum\alpha_k\cE^{(k)}_{a} = \sum \alpha_k \lb( \epsilon_{ab_1\ldots b_{d-1}}R^{b_1b_2}\wg\cdots\wg R^{b_{2k-1}b_{2k}} \wg e^{b_{2k+1}}\wg\cdots\wg e^{b_{d-1}} \rb) = 0.
        \ee
Written in terms of spacetime indices, we have the familiar terms
        \begin{subequations}\label{eq:known cE}
            \begin{align}
                \cE^{(0)}_{ij} & \sim \Lambda g_{ij} \\
                \cE^{(1)}_{ij} & \sim R_{ij} - \tfrac{1}{2}Rg_{ij}  = G_{ij} \\
                \cE^{(2)}_{ij} & \sim 2\lb( R_{iklm}R_j^{~klm} - 2 R_{ikjl}R^{kl} - 2 R_{ik}R^k_{~j} + RR_{ij} \rb) -\tfrac{1}{2} \cG g_{ij}.
            \end{align}
        \end{subequations}
The exact solutions describing black holes have been found for the Einstein-Gauss-Bonnet (EGB) as well as for the general Lovelock equations
(see \cite{boul deser}-\cite{whitt}). In what follows we shall consider the case of static spherically symmetric solutions in vacuum.\\

\section{Static Lovelock black holes}\label{sec:BH}

Let us consider the general static spherically symmetric metric to find the vacuum solution in LL-gravity. All vacuum spacetimes satisfy the null energy
condition, $\cG_{ab}k^ak^b = 0, \, k_ak^a = 0$ which means $\cG^t_t = \cG^r_r$ and that in turn implies $g_{tt}g_{rr} = const. = -1$. Thus the metric takes the form \cite{wheeler,whitt}:
        \be\label{eq:metric gen}
            ds^2 = - V(r)dt^2 +\frac{dr^2}{V(r)} + r^2 d\Omega_{d-2}^2
        \ee
    with
        \be\label{eq:V psi}
            V(r) = 1-\psi(r) r^2.
        \ee
The vacuum equation then reduces to solving the master algebraic equation,

        \be\label{eq:psi defn}
            \cF(\psi) = \sum \alpha_k \psi^k = \frac{\mu}{r^{d-1}}
        \ee
for $\psi(r)$. Here $\mu$ is the mass parameter of the solution. When $\mu=0$, Eq.\eqref{eq:psi defn} gives the vacua of the theory as the zeros
(say $\psi = \psi_*$) of the master polynomial $\cF(\psi)$ . Thus in general there will be as many vacua as there are zeros of the function $\cF(\psi)$.
These vacua can be anti-deSitter (AdS), Minkowski or deSitter (dS) depending on whether the particular $\psi_*$ is negative, zero or positive
respectively.\\

The vacua for which $\psi_*$ is not a simple zero behave differently. In this case, the field equations do not determine the gravitational
potential, $g_{tt}$ which remains completely arbitrary and free. Hence these were termed \emph{geometrically free} solutions by Wheeler
\cite{wheeler}. Such solutions correspond to the metric:
     \be\label{eq:metric free}
        ds^2 = - e^{2\phi}dt^2 +\frac{dr^2}{V(r)} + r^2 d\Omega_{d-2}^2
     \ee
where $\phi(r)$ is an undetermined arbitrary function and $V(r) = 1-\psi_0r^2$. Geometrically free vacua also arise in other situations in higher order
gravity theories \cite{dmk, BH soln}.\\

  The Lanczos-Lovelock Lagrangian Eq.\eqref{eq:lagrangian} has a large number of arbitrary dimensionful parameters given by the $N-1$ independent ratios
  of coupling constants $\alpha_k$. To reduce this arbitrariness in the theory, it becomes necessary to prescribe a relation among all the
  LL-coefficients. One  way is to restrict to the lowest order $\Lambda$ and the $N$th order;i.e. pure Lovelock gravity with $\Lambda$
  \cite{dad pure L}. The other is the case of the prescription of a unique $\Lambda$ giving the dimensionally continued black holes \cite{CTZ,BTZ}.  \\

    The general prescription for the Lovelock couplings to have a unique $\Lambda$ was given by Cris\'ostomo, Troncoso and Zanelli \cite{CTZ}. Their choice of LL-couplings is given by:
    \be\label{eq:CTZ alpha}
        \bar\alpha_k = \begin{cases}
                            \frac{1}{d-2k}\binom{K}{k}l^{2(k-K)}    & \text{for}\quad k \leq K  \\
                            0                                       & \text{for}\quad k > K
        \end{cases}
    \ee\\

    where $l$ is a length scale. In $d$ dimensions, this describes a family of gravity theories, labeled by the integer $K$ which represents the highest
    power of curvature that appears in the Lagrangian. Another prescription due to Ba\~{n}ados, Teitelboim and Zanelli \cite{BTZ} (hereafter referred to
    as \emph{BTZ-continuation}) is a special case of the above with $K = \lfloor \tfrac{d-1}{2}\rfloor = N$, taking its maximum possible value. The
    black hole solutions and their thermodynamic properties in these theories were discussed in \cite{CTZ,BTZ}. \\

    All these cases could be knit together in the degeneracy character of the master equation, Eq.\eqref{eq:psi defn}. The dimensionally
    continued black holes follow from the degeneracy of the master equation;viz

    \be\label{eq:CTZ F}
        \cF(\psi) = (\psi- \beta)^K
    \ee\\

while for the pure Lovelock black holes, it is

        \be\label{eq:cF pure Lovelock}
            \cF(\psi) = -\alpha + \psi^N
        \ee

    where $N$ is the maximum order of the Lovelock term which contributes in $d$-dimensions, given by Eq.\eqref{eq:N defn}. Note that $\alpha$ is up to
    a numerical factor the comological contant and we have set to unity the coefficient of the $N$th Lovelock term. Hence the solutions of Eq.
    (\ref{eq:psi defn}) will depend on the two parameters, $\alpha$ and $\mu$. Black hole solutions in such theories have been studied in
    \cite{dad pure L, pure LL}. Both these cases are synthesized in the generalization we consider which is the
    derivative degeneracy of the equation, and so we write

    \be\label{eq:cF dd}
        \cF(\psi) = -\alpha + (\psi- \beta)^N.
    \ee
The solution then takes the form

\be
V = 1 - \psi r^2 = 1 - \beta r^2 - r^2(\alpha + \frac{\mu}{r^{d-1}})^{1/N}.
\label{Vforlambdapurelovelockbeta}
\ee

Now $\alpha = 0$ gives the dimensionally continued black hole \cite{CTZ,BTZ} while $\beta = 0$ the pure Lovelock black hole \cite{dad pure L, pure LL}.
The latter therefore requires the
derivative degeneracy which in general combines the two. It is clear that the former cannot go to the Einstein solution asymptotically because there is
nothing to expand around while the latter has the correct Einstein limit because it can be expanded around $\alpha$ for large $r$. Then we readily get $V = 1 - ar^2 -b/r^{d-3}$ which is the $d$-dimensional Schwarzschild black hole in dS/AdS spacetime. The presence of $\alpha$ which acts
as the cosmological constant is therefore essential for existence of the correct Einstein limit asymptotically. It is the characterizing property of the
derivative degeneracy. At the high energy end for $r\to r_h$, it goes over to the dimensionally continued black hole solution while at the low energy
end for large $r$ to the corresponding Einstein solution. This is what is expected of the higher order terms in the Lagrangian that they should be
significant at the ultraviolet end while their effect should wean out at the infrared end. \\

Thus the derivative degeneracy generalization makes the black hole to have the right behavior at both high and low energy ends. For the pure Lovelock
black hole with $\beta=0$ in Eq. (\ref{Vforlambdapurelovelockbeta}), let us explicitly write the solution in odd and even dimensions as follows:

            \begin{subequations}\label{Vforlambdapurelovelock}
                \begin{align}
                   V &= 1 - \psi r^2 = 1 - (\alpha r^{2N}+ \mu)^{1/N} \ (d = 2N+1, \ {\rm odd})\\
                    V &= 1 - \psi r^2 = 1 - (\alpha r^{2N}+ \frac{\mu}{r})^{1/N}\ (d = 2(N+1), \ {\rm even}).
                \end{align}
            \end{subequations}

It is interesting to note that under the radical sign the potential due to the mass parameter $\mu$ is the same as that for the Einstein gravity in $3$
and $4$ dimensions. This indicates that the potential is essentially $N$th root of the Einstein potential in odd and even dimensions and it is this that makes the thermodynamics universal for the Lovelock black holes. That is what we take up in the next section. \\

There however remains the question of the unique fixing of the vacuum; i.e. the sign of $\alpha$. As has been argued in the appendix, $\alpha<0$
for the odd dimension while its sign for the even dimension is fixed by requiring the solution to go over to the corresponding Einstein solution
with positive mass. We would in the Sec.\ref{sec:asym} consider the asymptotic limit of the pure Lovelock black holes which would require mass,
$\bar\mu$ as given below in Eq.\eqref{eq:effective param pure L} to be positive. That fixes for the even dimension $\alpha > 0$ uniquely  (for $N > 1$). Thus there remains no more ambiguity about the unique realization of the vacuum. \\

\section{Thermodynamics and the Characterization of the pure Lovelock black holes}\label{sec:therm}

The causal structure and thermodynamical properties of general spherically symmetric solutions in LL-gravity have been extensively studied
(\cite{whitt, cai-therm}). The temperature and entropy can be easily computed using the methods given in \cite{jac-my,cai-therm} and so we have

\begin{subequations}
    \begin{align}
        \begin{split}\label{eq:dd temp}
            T & = \frac{1}{4\pi} V'(r_h ) \\
              & = -\frac{1}{4\pi} \lb( \frac{2}{r_h}+r_h^2\psi'(r_h) \rb) \\[15pt]
        \end{split}\\
        S & = \int T^{-1}d\mu = \int_0^{r_h} T^{-1}\frac{\partial \mu}{\partial r_h} dr_h \label{eq:dd entropy}
    \end{align}
\end{subequations}

where $r_h$ is the radius of the horizon. The entropy $S$ is in general a polynomial series given in \cite{cai-therm} and does not simplify.

In the case of BTZ-continuation ($\alpha = 0$), the temperature and entropy can be written as:

\begin{subequations}
    \begin{align}\label{eq:BTZ temp-entropy}
        T   = \begin{cases}
               \frac{\mu^{2/(d-1)} -1}{2\pi r_h} & d = odd  \\[10pt]
              \frac{1}{2\pi}\lb[ - \frac{1}{r_h} + \frac{d-1}{d-2}~\frac{\mu^{2/(d-2)}}{r_h^{d/(d-2)}}\rb] & d = even   \\
          \end{cases}\\[15pt]
        S  = \begin{cases}
                    2\pi \lb[(d-1)\mu^{(d-3)/(d-1)} \rb] ~ r_h & d=odd  \\
                    2\pi \lb[\frac{(d-2)^2}{d}\mu^{(d-4)/(d-2)}\rb] ~ r_h^{d/(d-2)} & d = even  \\
            \end{cases}
    \end{align}
\end{subequations}

Note that in the above the temperature and entropy in the odd dimension have the universal relation to the horizon radius but not so for the
even dimension. In the odd dimension, gravitational potential due to mass is constant while that due to the cosmological constant has the
universal $r$ dependence and that is what defines the black hole in this case. Thus it is no surprise that thermodynamics has the universal
character in odd dimension. In contrast the BTZ continuation black holes in even dimension where mass also plays active role with potential
being $r$ dependent, they do not have universal thermodynamical behavior as is clear from the above expressions for temperature and entropy. We
shall now show the universality for the pure Lovelock black holes. \\

In the case of pure-Lovelock where $\beta=0$, the thermodynamical quantities are given by

\begin{subequations}
    \begin{align}\label{eq:LL T}
        T   = \begin{cases}
               \frac{\mu -1}{2\pi r_h} & d = odd    \\[10pt]
              \frac{1}{2\pi}\lb[- \frac{1}{r_h} + \frac{d-1}{d-2}~\frac{\mu}{r_h^2}\rb] & d = even  \\
          \end{cases}\\[15pt]
            \label{eq:LL S}
        S  = \begin{cases}
                    2\pi (d-1) ~r_h & d=odd \\
                    \pi (d-2) ~r_h^2 & d = even \\
            \end{cases}
    \end{align}\label{TandS}
\end{subequations}

This demonstrates the universal thermodynamical behavior in terms of the event horizon radius. That is temperature and entropy always bear the
same relation to the horizon radius. This is the remarkable property of the pure Lovelock black holes \cite{dad pure L}. ( In odd
dimension $\alpha$ has to be negative for positive $\mu-1$. So was the case for the famous BTZ black hole in $(2+1)$ dimension \cite{BTZ0} where
$V(r) = r^2/l^2 - \mu$ which clearly indicated negative $\Lambda$. Thus in odd dimension the cosmological constant has always to be
negative;i.e. an AdS. In even dimension, the condition for the temperature to be positive is guaranteed by Eq. (\ref{restr-dS-2}). ) Not only
that the converse is also true, the universality uniquely characterizes the Lovelock black holes. That is what we show next.

Using the general formulae obtained in \cite{cai-therm}, it
can be easily shown that pure-Lovelock theory is the unique one which has such universal thermodynamical behaviour. Explicity these are given by a
series in terms of powers of the horizon radius $r_h$ as  ($\Omega_d= \frac{2\pi^{\frac{d-2}{2}}}{\Gamma(\frac{d-2}{2})} $):

\begin{subequations}
    \begin{align}
        T & = \frac{\sum_{i=0}^N (d-2i-1)\alpha_ir_h^{-2i+2}}{4\pi r_h \sum_{i=1}^{N}i\alpha_i r_h^{-2i+2}}  \\
        S & = \frac{\Omega_d r_h^{d-2}}{4G} \sum_{i=1}^N\frac{i(d-2)}{d-2i} \alpha_ir_h^{-2i+2}
    \end{align}
\end{subequations}

 For universality of thermodynamics we would now demand that the temperature and entropy are always given in terms of the horizon radius as for the
 Einstein gravity in $3$ and $4$ dimensions. That is their horizon radius dependence is entirely free of the spacetime dimension and the Lovelock order.
 This means in the series for the black hole entropy $S$, for $\alpha_i \neq 0$ we must have
 \be
    r_h^{d-2i} =    \begin{cases}
                r_h & d=odd \\
                r_h^2 & d=even
            \end{cases}
\ee
Thus $i = \lb\lfloor\frac{d-1}{2}\rb\rfloor = N$ and so the only terms that
contribute are $\alpha_0$ and $\alpha_N$; i.e. $\Lambda$ and the maximal order Lovelock. This is what characterizes the pure Lovelock black hole. This
proves the sufficient condition that the universality uniquely singles out the pure Lovelock gravity. \\

It should however be noted that this universality is exhibited when we express thermodynamics in terms of mass $\mu$ and the radius $r_h$, which are the
natural black hole  parameters. Instead had we written it all in terms of $\mu$ and $\alpha$, it would have been missed. It is therefore important to tag
on the right black hole parameters to study its thermodynamics. \\

Further like the BTZ-continuation black holes, the Einstein-Gauss-Bonnet black hole also does not have this universal behavior. The temperature
and entropy for the Einstein-Gauss-Bonnet theory, $\cF(\psi) = -\alpha + \psi + \gamma \psi^2$, read as follows:

\begin{subequations}
    \begin{align}\label{eq:EGB temp-entropy}
        T   =  \frac{1}{2\pi}\lb[- \frac{1}{r_h} + \frac{d-1}{2}~\frac{\mu}{r_h^{d-2} \lb( 1+\tfrac{2\gamma}{r_h^2} \rb)}\rb] \\[15pt]
        S  = \frac{4\pi}{d-2}\lb[ 1+\frac{2 (d-2)\gamma}{r_h^2} \rb] r_h^{d-2}
    \end{align}
\end{subequations}
where $\gamma$ is the Gauss-Bonnet coupling parameter. Clearly thermodynamics in terms of the horizon radius is not universal. Here we have considered the $N=2$ Gauss-Bonnet case but the same would be the case for any order $N$ which means the Einstein-Lovelock black holes will not in general respect the thermodynamical universality.

Thus universality is the unique characterizing property of the pure Lovelock black holes. It is both necessary and sufficient condition.

\section{Asymptotic behavior}\label{sec:asym}

    When the spacetime contains a mass source $\mu \neq 0$, near spatial infinity the spacetime tends to one of its vacua $\psi_*$
    (from Eq.\eqref{eq:psi defn}). If $\cF(\psi)$ has no real zero, then the spacetime does not contain any spatial infinity. The behaviour far away
    from the mass source, or equivalently, near spatial infinity, is given by the Taylor expansion of $V(r)$ around the particular vacuum
    $\psi = \psi_*$, (see \cite{whitt}):

        \be\label{eq:V gen expand}
            V(r) \approx 1 - \psi_* r^2 - \frac{1}{r^{d-3}} \lb( \frac{\mu}{\cF'_*}\rb) + \ldots
        \ee

        where $\cF'_* = \lb.\frac{d\cF}{d\psi}\rb|_{\psi=\psi_*}$.\\

    Note that this is the asymptotic limit only when $\cF'_*\neq 0$, i.e. $\psi_*$ a simple zero of $\cF$. It represents a $d$-dimensional  Einstein
    black hole of mass $\bar \mu = \frac{\mu}{\cF'_*}$ in the deSitter spacetime with $\bar\Lambda = \tfrac{1}{2}\psi_*$. Consequently, near spatial
    infinity corresponding to a simple zero, every static spherically symmetric LL-spacetime with the cosmological constant behaves like the familiar
    Einstein one.\\

If $\psi_*$ is a degenrate zero of $K^{th}$ order, the expansion has to be carried down to this order,
        \be\label{eq:expand cF}
            \cF(\psi) = \cF_* + \cF'_* (\psi - \psi_*) + \frac{1}{2!}\cF''_* (\psi-\psi_*)^2 + \ldots
        \ee

and the asymptotic limit would then be given by

        \be\label{eq:V free expand}
            V(r) \approx 1 - \psi_* r^2 - \frac{1}{r^{\tfrac{d-1}{K}-2}} \lb( \frac{K!~\mu}{\cF^{(K)}_* } \rb)^\frac{1}{K}
        \ee
Clearly, this is not the Einstein limit unless $K=1$. Thus, general Lanczos-Lovelock spacetimes will not have asymptotic Einstein limit when the
polynomial has degenerate zeros. 

There could however be the case of both a simple zero and degenerate zero, for example

    \be\label{eq:cF factor}
        \cF(\psi) = (\psi - \psi_0)(\psi - \psi_1)^2
    \ee
    In this case, $\psi_* = \psi_0$ is a simple zero, but $\psi_* = \psi_1$ is doubly-degenerate. The $\mu \neq 0$ vacuum solution will then have two
    spatial infinities corresponding to each $\psi_0$ and $\psi_1$. Expansion near the infinity corresponding to $\psi_0$ will show  an Einstein-like
    behavior (as in Eq.\eqref{eq:V gen expand}), but that near $\psi_1$ will not.\\[15pt]

    The special case of derivative degenerate theory given by Eq.\eqref{eq:cF dd} will give vacua at $\psi_* = \beta + \alpha^{\tfrac{1}{N}}$. These will
    correspond to a simple zero iff $\alpha\neq 0$ and hence will always have Einstein-like limit. The case $\alpha = 0$ corresponds to BTZ-continuation where there is a unique spatial infinity $\psi_* = \beta$. This spatial infinity always
    corresponds to a degenerate zero unless $N=1$ (viz. Einstein gravity in $d=3,4$). Thus degeneracy of the master polynomial \eqref{eq:CTZ F} defines a unique spatial infinity but these solutions will not have the Einstein-like asymptotic behavior unless
    $N=1$ which is anyway the Einstein gravity. \\

    For pure-Lovelock gravity, Eq.\eqref{eq:cF pure Lovelock} gives $\psi_* = \lb(\alpha\rb)^{\tfrac{1}{N}}$ and $\cF'_* = N \alpha^{\tfrac{N-1}{N}}$.
    Thus, $\cF$ has a single simple zero if and only if $\Lambda \neq 0$, and all $\Lambda$ + pure-Lovelock theories will have an asymptotic limit
    similar to the Einstein theory with the effective values:

            \begin{subequations}\label{eq:effective param pure L}
                \begin{align}
                \bar \mu        & = \frac{\mu}{N \alpha^{\tfrac{N-1}{N}}} \label{eq:m effective}    \\
                    \bar\Lambda     & = \tfrac{1}{2}\alpha^{\tfrac{1}{N}}   \label{eq:Lambda effective}
                \end{align}
            \end{subequations}

        Thus for all orders in the Lovelock polynomial, the asymptotic limit of the pure-Lovelock solution, is the same as the Einstein solution with
        effective cosmological constant and mass given by Eq.\eqref{eq:effective param pure L}. This generalizes the analysis and bears out the
        expectation of Dadhich \cite{dad pure L} that in presence of a cosmological constant the spherically symmetric vacuum solution in any theory
        whether strictly pure-Lovelock, $\Lambda$ + only $N$th order,  or Einstein-Lovelock, $\sum \alpha_kL^k$, always goes over to the
        corresponding Einstein solution asymptotically. This is a universal asymptotic behavior of the LL  vacuum solutions. \\

    This general analysis shows why the cosmological constant is needed for this to work. If $\bar\Lambda = 0$, then any order $N > 1$ pure-Lovelock
    gravity will have a degenerate zero at $\psi_* = 0$. The solution (from Eq.\eqref{eq:V free expand}), has non-Einstein-like asymptotic behavior:

        \be\label{eq:V expand free pure L}
            V(r) \approx 1 -  \frac{(N!\mu)^\frac{1}{N}}{ r^{\tfrac{d-1}{N}-2}}
        \ee\\

    Thus the presence of a non-zero cosmological constant in a pure-Lovelock theory,  prevents the occurrence of degenerate zeros and hence all
    pure-Lovelock theories have asymptotic behavior like the Einstein theory.\\

\section*{Conclusion}

 There had been very extensive and rich literature on the Lovelock black holes and hence it is pertinent to ask what is it that this paper does which is
 new and non-trivial? First of all it synthesizes the dimensionally continued and the pure Lovelock black holes into the derivative degeneracy property
 of the master equation (9). The characterizing feature of this property is that it renders the black hole in contrast to the BTZ continuation have the right
 limits at both ends, $r\to r_h$ and large $r$. This is what it should be from  physical point of view because the higher order curvature effects should
 wean out at low energy. For the latter limit, the presence of $\alpha$ in  Eq. (14) is essential which marks the derivative degeneracy so that the
 potential could be expanded around it. The realization that at the root of these two family of solutions is simply the degeneracy character of the
 master equation is new and rather novel. The pure Lovelock black holes or the black holes having proper Einstein limit asymptotically could be viewed
 as physical realization of the the derivative degeneracy condition. \\

The most remarkable feature of the pure Lovelock black holes is their thermodynamical universality which means the temperature and entropy bear
the same relation to the event  horizon radius as the Einstein black hole in odd $3$ and even $4$ dimension irrespective of the Lovelock order
$N$. And the universality of thermodynamics uniquely identifies the pure Lovelock black holes \cite{kpd-lett}. Thus the necessary and sufficient
condition for the thermodynamical universality of a black hole is that it be a pure Lovelock black hole. Further it is interesting to note that
the potential inside the radical in Eq. (16) for odd/even dimension has the same form as the Einstein potential for $3/4$ dimension. This is
what makes the thermodynamics universal and more importantly it signifies that the gravity in $d=2N+1,\, 2(N+ 1)$ dimensions for the
$N$th order Lovelock has a kind of universal character. This is a remarkable realization which has been established firmly and explicitly in \cite{odd}. \\

The universality of the asymptotic behavior imitating the Einstein gravity as it was envisaged in \cite{dad pure L} has been established in general for
the pure Lovelock as well as for the Einstein-Lovelock for any order. For the latter it was verified earlier only upto $N=3$ while our analysis extends
this to all orders. It is remarkable that the derivative degeneracy beautifully imbibes the asymptotic Einstein limit and also brings out the essential
role of the non-zero $\Lambda$. 

Further very recently the universality of the Lovelock vacuum in the odd critical dimension has been established \cite{odd}. As vacuum is trivial in $3$ dimension for the Einstein gravity, similarly the same is also true for higher order pure Lovelock gravity for the corresponding higher order Riemann analogue. That is, the static vacuum solution in $(d=2n+1)$ dimension has vanishing $n$th order analogue of Riemann. However it is not Riemann flat.\\

We conclude by reiterating that the derivative degeneracy leads to an interesting synthesis of the BTZ continuation and pure Lovelock black
holes and it renders the black hole to have proper right limits at both the ends, small and large $r$. The realization of the thermodynamical
universality of pure Lovelock black holes is very enlightening and insightful for understanding gravitational dynamics in higher dimensions. It
would be interesting to see whether the universality bears out even when entropy is computed by quantum calculations. We believe that it should
hold true because it is computed by integrating the First Law of black hole thermodynamics which should transcend to the quantum computations as
well.

\section*{Appendix}\label{sec:app}

Here we study some features of eq. (\ref{Vforlambdapurelovelock}) for $d$ even ($d=2(N+1)$) and odd ($d=2N+1$), with $\mu>0$.

\subsection{$\alpha>0$\,,\ $d$  even}

For the background to have physical sense, $V$ must be positive at the maximum $\displaystyle r_0= (\frac{\mu}{2 N \alpha})^{\frac{1}{1+2 N}}$, which
implies for given $\alpha$ an upper limit for the mass of the BH,
\be
\mu^{2 N} \alpha < \frac{(2N)^{2N}}{(2N+1)^{2N+1}}
\label{restr-dS}
\ee
The horizon of the BH is located at a value for the radial coordinate, $0<r_h<r_0$ for which $V(r_h)=0$.
In terms of the BH parameters, the allowed values are
 \be
0<r_h-\mu< \frac{\mu}{2N}\,.
\label{restr-dS-2}
\ee
From $r_0$ onwards, $V$ decreases until it reaches again the zero value ar $r=r_s>r_0$, which is a coordinate singularity similar to the one present in
the de Sitter background in static coordinates (see for instance \cite{BD}).

\subsection{$\alpha>0$\,,\ $d$  odd}

To makes sense of the metric, we need $\mu<1$. There is only the coordinate singularity at $\displaystyle r_s= (\frac{1-\mu}{ \alpha})^{\frac{1}{2 N}}$
but no BH horizon. Notice from equation (\ref{TandS}) that precisely $\mu>1$ is required for a positive temperature. We conclude that this case must be
excluded.

\subsection{$\alpha<0$\,,\  $d$  even}

We write $\alpha=-l^2$.
For $r$ small, $r\to 0$, $V$ is large and negative. It increases with increasing $r$ and its vanishing defines the BH horizon. From
$\displaystyle r_1 = (\frac{\mu}{l^2})^\frac{1}{2 N+1}$ onwards we will need $N$ odd to keep $V$ real. Thus $d$ is $d= 2(N+1)=\dot 4$.

\subsection{$\alpha<0$\,,\  $d$ odd}

For small $r$ the quantity $\displaystyle (-l^2\, r^{2N}+\mu)$ is positive. If $\mu<1$ there will be no BH horizon. If $\mu>1$ there is a BH horizon at
$\displaystyle r_h= (\frac{\mu-1}{l^2 })^{\frac{1}{2 N}}$. From $\displaystyle r_1 = (\frac{\mu}{l^2})^\frac{1}{2 N}$ onwards we will need $N$ odd to
keep $V$ real. This means that $d= 2N+1 =2(N+1)-1 =\dot 4 +3$.

Note from the analysis above that for $\alpha<0$ one always needs $N$ to be odd, in agreement with previous findings in section \ref{sec:asym}
(see \eqref{eq:effective param pure L}).

\end{document}